\documentclass{PoS}

\usepackage{amsmath}
\usepackage[normalem]{ulem}

\usepackage{enumitem}

\title{Cosmic Neutrinos and the Cosmic-Ray Accelerator TXS 0506+056}

\ShortTitle{On the Origin of Cosmic Neutrinos}

\author{\speaker{Francis Halzen}\\
Wisconsin IceCube Particle Physics Center\\
University of Wisconsin{\textendash}Madison, Madison, WI USA\\
        E-mail: \email{halzen@icecube.wisc.edu}}

\abstract{

IceCube discovered a flux of cosmic neutrinos originating in extragalactic sources with an energy density close to that in gamma rays and cosmic rays. A multimessenger campaign triggered by the coincident observation of a gamma-ray flare and a 290-TeV IceCube neutrino pinpointed the cosmic-ray accelerator TXS 0506+056. Subsequently, the IceCube archival data revealed a 3-month burst of 13 cosmic neutrinos in 2014-15 that dominates the neutrino flux of the source over the 9.5 years of observations.  The original identification of the source as a blazar was puzzling because it requires a major accretion event onto the rotating supermassive black hole to accommodate the neutrino burst. Subsequent high-resolution radio images of the source with the VLBA brought to light a merger of two galaxies, revealed by the interaction of two jets entangled in the source. Recently, the blazar PKS 1502+106 was found in the direction of a 300-TeV neutrino alert, IC-190730. OVRA radio observations at 15\,GHz indicate that the neutrino also coincides with the highest flux density of a flare that started five years ago. This matches the similar long-term outburst seen from TXS~0506+056 and may indicate merger activity. Also, the dominant hotspot in the 10-year IceCube neutrino sky map, NGC 1068 (Messier 77), is a Seyfert galaxy undergoing a major accretion event onto the black hole. A few-percent fraction of such special sources, now labeled as gamma-ray blazars, is sufficient to accommodate the diffuse cosmic neutrino flux observed by IceCube. While rapid progress seems likely, the observations also convincingly make the case for the construction of more and larger neutrino telescopes with better angular resolution.\\

\vspace{4mm}
{\bfseries Corresponding authors:}
{Francis Halzen}$^{1}$\\
{$^{1}$ \itshape UW--Madison}
}

\FullConference{36th International Cosmic Ray Conference-ICRC2019-\\
		July 24th - August 1st, 2019\\
		Madison, WI, U.S.A.}

\begin{document}

\section{Detecting cosmic neutrinos}\label{sec1}
The highest energy radiation reaching us from the universe is not radiation at all; it is cosmic rays---high-energy nuclei, mostly protons.
We do not know where they come from or how they are accelerated to extreme energies exceeding $10^8$~TeV. However, the identification of  a distant rotating supermassive black hole as the first cosmic-ray accelerator has provided a breakthrough indicating a promising path towards the solution of the cosmic-ray puzzle: multimessenger astronomy~\cite{IceCube:2018dnn,IceCube:2018cha}. 

The rationale for searching for cosmic-ray sources by observing neutrinos is straightforward: in the particle flows onto neutron stars or black holes, some of the gravitational energy released in the accretion of matter is transformed into the acceleration of protons or heavier nuclei, which subsequently interact with ambient radiation, gas, and molecular clouds to produce astrophysical neutrinos originating from the decay of pions and other secondary particles. Both neutral and charged secondary pions are produced, and while charged pions decay into neutrinos, neutral pions decay into gamma rays. The fact that cosmic neutrinos are inevitably accompanied by high-energy photons transforms neutrino astronomy into multimessenger astronomy.  A challenge of multimessenger astronomy is to separate these photons, referred to as {\em pionic photons}, from photons radiated by electrons that may be accelerated along with the cosmic rays.

The search for high-energy astrophysical neutrinos led to the development of large scale neutrino detectors: DUMAND in the Pacific Ocean, NT200 in Lake Baikal, AMANDA at the South Pole, and ANTARES in the Mediterranean Sea. These detectors instrumented transparent natural water or ice with photomultipliers to build Cherenkov detectors; see \cite{Katz:2011ke}. The pattern of Cherenkov light radiated by secondary charged particles produced in interactions of neutrinos inside or near the instrumented volume encode the flavor, energy, and arrival direction of the neutrino. Based on the anticipation that cosmic accelerators may produce similar energies in cosmic rays, gamma rays, and neutrinos, it had been anticipated that the construction of a cubic-kilometer, gigaton-scale detector would be sufficient to make a first detection of high-energy neutrinos of cosmic origin~\cite{Gaisser1995}.

The IceCube Neutrino Observatory~\cite{Aartsen:2016nxy} transformed one cubic kilometer of natural Antarctic ice below the National Science Foundation's research station at the geographical South Pole into a Cherenkov detector. Constructed between 2004 and 2010, it has now taken 10 years of data with the completed detector. Below a depth of 1450\,meters, a cubic-kilometer of glacial ice is instrumented with 86 cables called ``strings,'' each of which is equipped with 60~digital optical modules (DOMs). The DOM is a glass pressure vessel containing a 10-inch photomultiplier tube (PMT) and digitizing electronics that capture and time stamp the PMT light signal~\cite{Aartsen:2016nxy}. These digitized waveforms are the basis of IceCube analysis: the recorded arrival time and amplitude of the photons are used to reconstruct the direction and energy of neutrinos of different flavors interacting in the ice.

\begin{figure}[t]\centering
\includegraphics[width=0.43\linewidth]{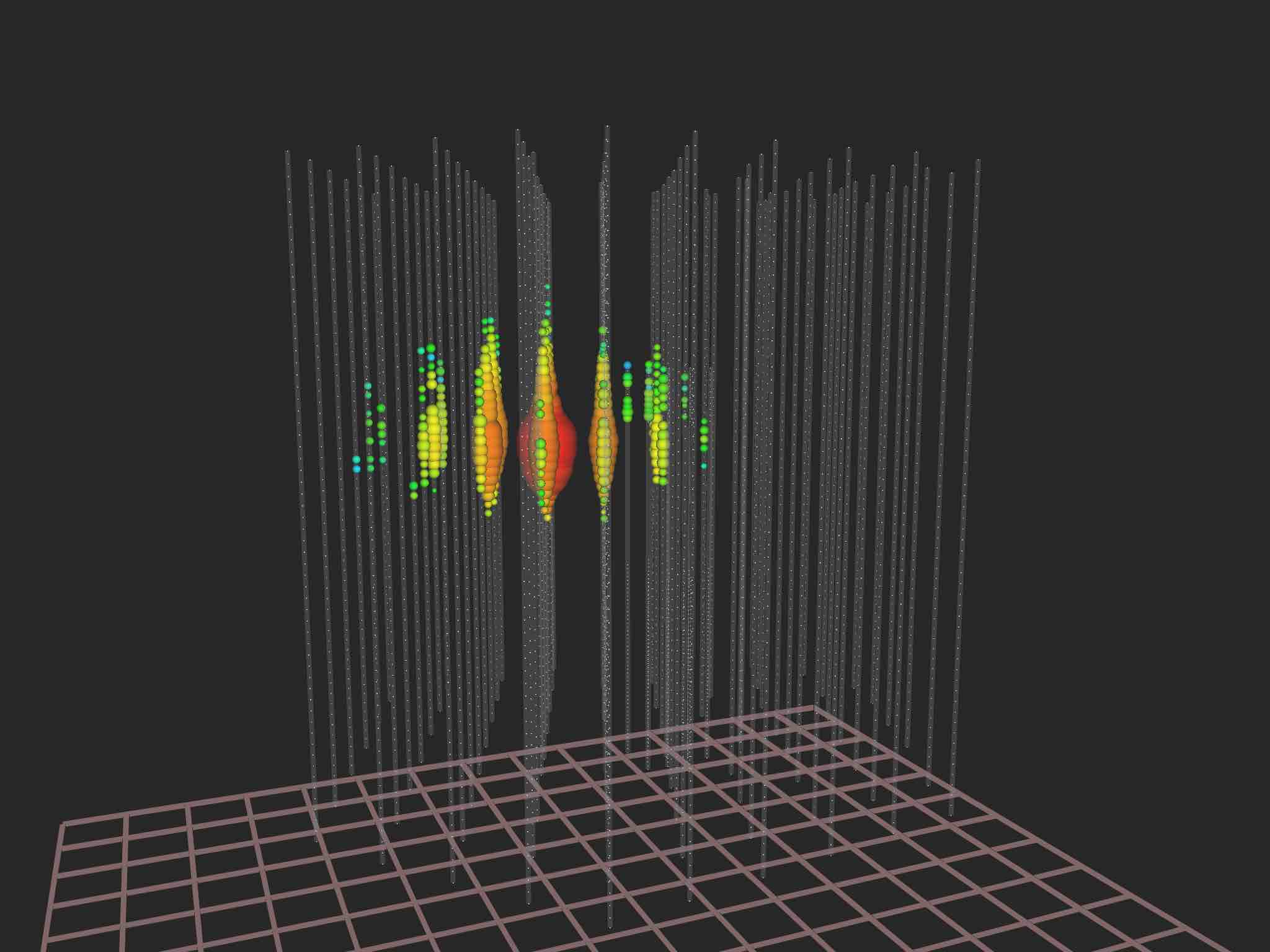}\hspace{0.5cm}\includegraphics[width=0.43\linewidth]{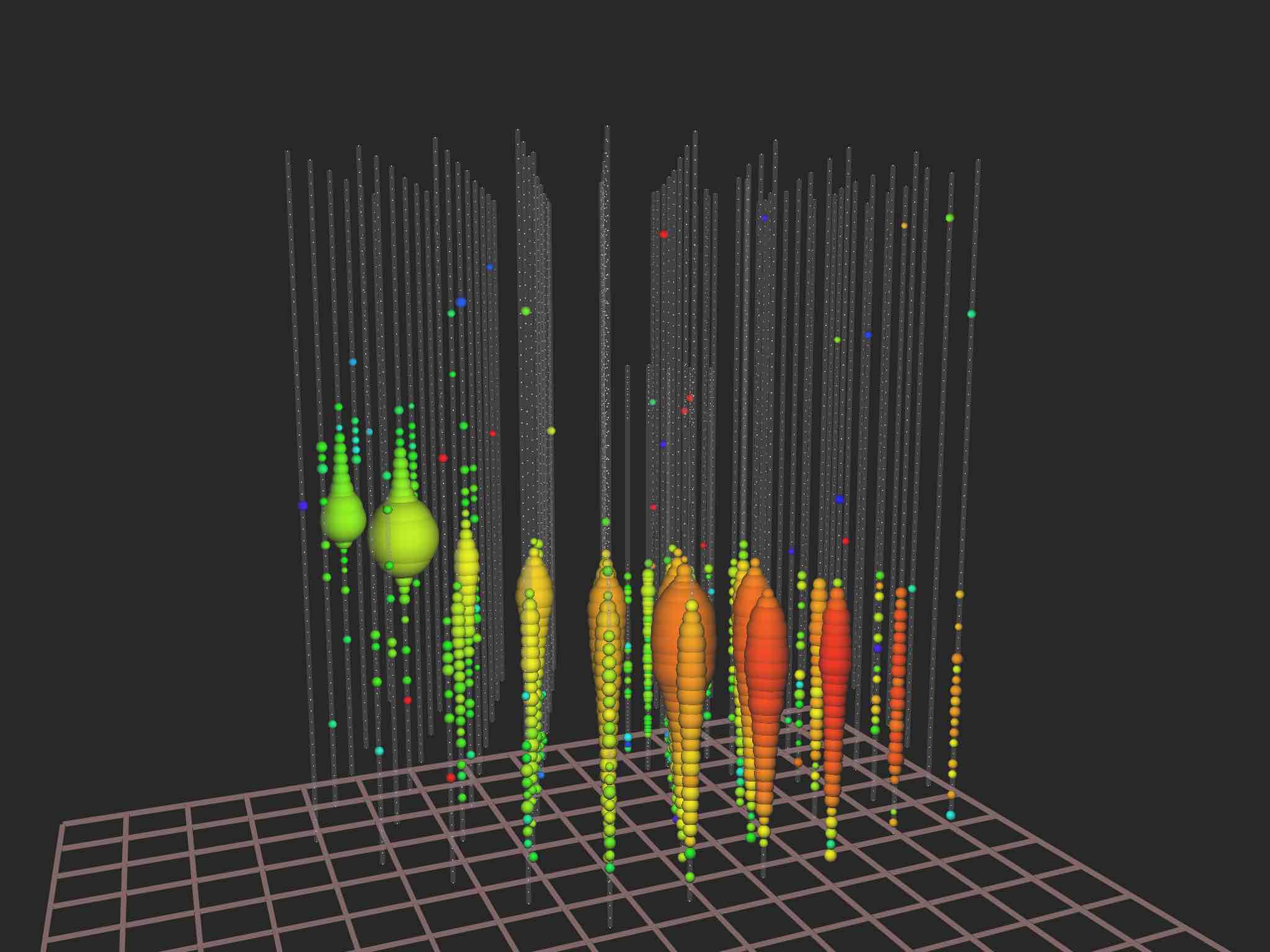}
\caption[]{{\bf Left Panel:}  Light pool produced in IceCube by a shower initiated by an electron or tau neutrino of $1.14$ PeV, which represents a lower limit on the energy of the neutrino that initiated the shower. White dots represent sensors with no signal. For the colored dots, the color indicates arrival time, from red (early) to purple (late) following the rainbow, and size reflects the number of photons detected. {\bf Right Panel:}  A muon track coming up through the Earth, traverses the detector at an angle of $11^\circ$ below the horizon. The deposited energy, i.e., the energy equivalent of the total Cherenkov light of all charged secondary particles inside the detector, is 2.6\,PeV.}
\label{fig:erniekloppo}
\end{figure}
The direction of a muon track or electromagnetic shower initiated by an electron or tau neutrino is determined by the arrival times of photons at the optical sensors, while the number of photons is a proxy for the energy deposited by secondary particles in the detector. For illustration, the Cherenkov patterns initiated by an electron (or tau) neutrino of 1\,PeV energy and a neutrino-induced muon losing 2.6\,PeV energy while traversing the detector are contrasted in Fig.~\ref{fig:erniekloppo}. For the cascade (shower) event shown in the left panel of Fig.~\ref{fig:erniekloppo}, more than 300 digital optical modules report a total of more than 100,000 photoelectrons. 

Long tracks resulting from muon neutrino interactions can be pointed back to their sources with a $\le 0.4^\circ$ angular resolution.  In contrast, the reconstruction of cascade directions, in principle possible to a few degrees, is still in the development stage in IceCube~\cite{Aartsen:2013vja, Tyuan2017}. Determining the deposited energy from the observed light pool is, however, relatively straightforward, and a resolution of better than 15\% can be achieved.  
 
In its search for high-energy neutrinos of cosmic origin, IceCube continuously monitors the whole sky. The instrumentation thus collects very high statistics data sets of cosmic ray muons and atmospheric neutrinos. Neutrino energies cover more than six orders of magnitude, from 5~GeV in the highly instrumented inner core (DeepCore) to beyond 10~PeV. Soon after the completion of the detector, with two years of data, IceCube discovered an extragalactic flux of cosmic neutrinos with an energy density in the local universe that is similar to that in gamma rays; see~\cite{Ackermann:2014usa, Fang:2017zjf}.

\begin{figure}[ht!]
\centering
\includegraphics[width=0.95\linewidth]{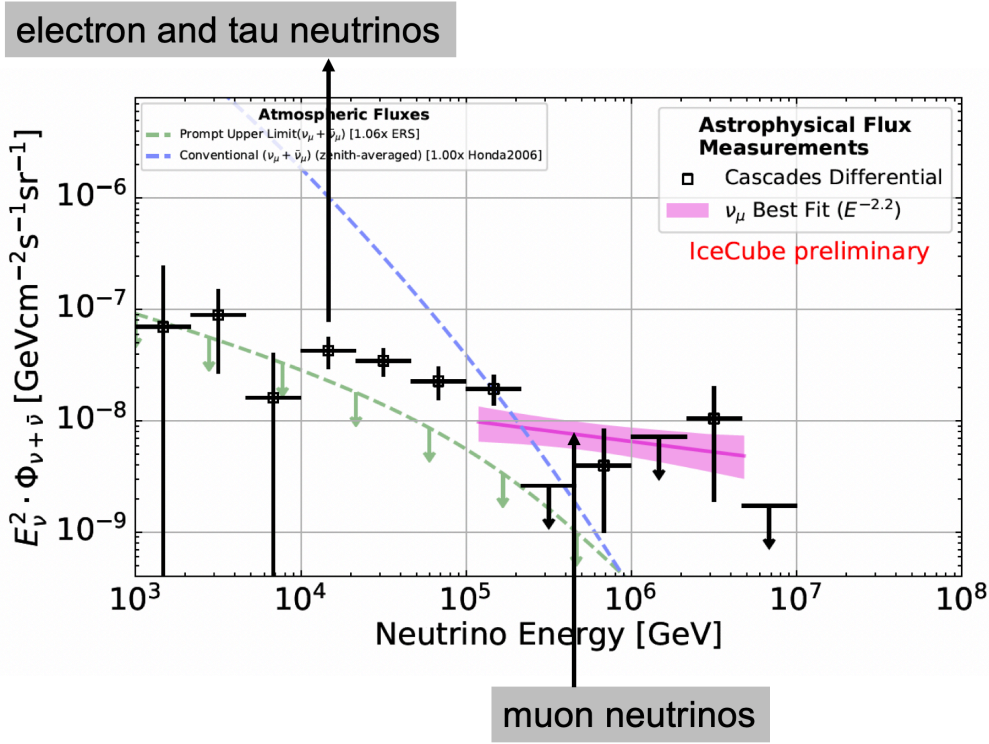}
\caption{The spectral flux ($\phi$) of cosmic muon neutrinos inferred from the eight-year upgoing muon track analysis (red solid line) with $1\sigma$ uncertainty range (shaded range) is compared with the flux of showers initiated by electron and tau neutrinos~\cite{Aartsen:2017mau}. The measurements are consistent assuming that each flavor contributes the same flux to the diffuse spectrum.}
\label{fig:showerstracks}
\end{figure}

Two principal methods are used to identify cosmic neutrinos. The first method reconstructs upgoing muon tracks initiated by muon neutrinos. The kilometer-long muon range makes it possible to identify neutrinos that interact outside the detector and to separate them from the the background of atmospheric muons using Earth as a filter. With this method, IceCube has measured the background atmospheric neutrino flux over more than five orders of magnitude in energy with a result that is consistent with theoretical calculations. However, with eight years of data, IceCube has observed an excess of neutrino events at energies beyond 100\,TeV~\cite{Aartsen:2015rwa,Aartsen:2016xlq,Aartsen:2017mau} that cannot be accounted for by the atmospheric flux. Although the detector only records the energy of the secondary muon inside the detector, from Standard Model physics we can infer the energy spectrum of the parent neutrino. The high-energy cosmic muon neutrino flux is well described by a power law with a spectral index of $2.19\pm0.10$ and a normalization at 100\,TeV neutrino energy of $(1.01^{+0.26}_{-0.23})\,\times10^{-18}\,\rm GeV^{-1}\rm cm^{-2} \rm sr^{-1}$~\cite{Aartsen:2017mau}.

The second method exclusively identifies high-energy neutrinos interacting inside the detector, so-called high-energy starting events. It divides the instrumented volume of ice into an outer veto shield and a $\sim500$-megaton inner fiducial volume. The advantage of focusing on neutrinos interacting inside the instrumented volume of ice is that the detector functions as a total absorption calorimeter~\cite{Aartsen:2013vja} allowing for a good energy measurement that separates cosmic from lower-energy atmospheric neutrinos. Additionally, atmospheric neutrinos reaching us from the Southern Hemisphere are readily distinguishable because they are accompanied by particles produced in the same air shower where the neutrinos originate. With this method, neutrinos from all directions in the sky and of all flavors can be identified, which includes both muon tracks as well as secondary showers produced by charged-current interactions of electron and tau neutrinos and neutral current interactions of neutrinos of all flavors. A sample event with a light pool of roughly one hundred thousand photoelectrons extending over more than 500 meters is shown in the left panel of Fig.~\ref{fig:erniekloppo}.

Both measurements yield consistent determinations of the cosmic neutrino flux; see Fig.~\ref{fig:showerstracks}. It was the HESE method that succeeded in revealing the first evidence for neutrinos of cosmic origin~\cite{Aartsen:2013bka,Aartsen:2013jdh}. Events with PeV energies, and no trace of accompanying muons from an atmospheric shower, are highly unlikely to be of atmospheric origin. The seven-year data set contains a total of 60 neutrino events with deposited energies ranging from 60\,TeV to 10\,PeV. The data are consistent with an astrophysical component with a spectrum close to $E^{-2.2}$ above an energy of $\sim 200$\,TeV~\cite{Aartsen:2017mau}.

We should mention here that there is yet another method to conclusively identify cosmic neutrinos: the observation of very high energy tau neutrinos. Tau neutrinos produce two spatially separated showers in the detector, one from the interaction of the tau neutrino and the second one from the tau decay; the mean tau lepton decay length is about 50~m/PeV. Two such candidate events have been recently identified~\cite{ignationeutrino2018}. In addition to the double-cascade candidates, we should also mention that a first candidate event has been attributed to the Glashow resonance. This event was identified in a search for partially contained events: an anti-electron neutrino interacting with an atomic electron produced an event with an energy of 6.3 PeV, characteristic of the resonant production of a weak intermediate W boson~\cite{Aartsen:2017mau, lulu_UHECR}. 

\section{IceCube neutrinos and Fermi photons}\label{IC_Fermi}

The most important message emerging from the IceCube measurements of the high-energy cosmic neutrino flux is not apparent yet: the prominent role of hadrons relative to leptons in the high-energy universe. As highlighted in the introduction, photons are inevitably produced in association with neutrinos when accelerated cosmic rays produce neutral and charged pions in interactions with target material in the vicinity of the accelerator. Targets include strong radiation fields that may be associated with the collapsed objects that power the accelerator, as well as concentrations of matter, such as hydrogen or molecular clouds in their vicinity.
The relative production rates of pionic gamma rays and neutrinos only depend on the ratio of charged-to-neutral pions produced in cosmic-ray interactions. For interactions of protons with a gamma ray target, the production of pions is dominated by the $\Delta$ resonance: $p + \gamma \rightarrow \Delta^+ \rightarrow \pi^0 + p$ and $p + \gamma \rightarrow \Delta^+ \rightarrow \pi^+ + n$. These channels produce charged and neutral pions with probabilities of 2/3 and 1/3, respectively. In contrast, if neutrinos originate in the interactions of a hadronic beam with matter, e.g.,~hydrogen or molecular clouds, equal numbers of pions of all three charges emerge: $p+p \rightarrow N_\pi\,[\,\pi^{0}+\pi^{+} +\pi^{-}]+X$, where $N_\pi$ is the pion multiplicity. 
The production of pionic neutrinos and gamma rays can be related without any reference to the cosmic-ray beam that initiates their production in the target; the relation simply reflects the fact that while neutral pions decay as $\pi^0\to\gamma+\gamma$ and create a flux of high-energy gamma rays, the charged pions decay into three high-energy neutrinos ($\nu$) and antineutrinos ($\bar\nu$) via the decay chain $\pi^+\to\mu^++\nu_\mu$ followed by $\mu^+\to e^++\bar\nu_\mu +\nu_e$ and the charged-conjugate process. 

Before applying this relation to data, one must realize that, unlike neutrinos, the universe is not transparent to PeV gamma rays. They will interact and initiate an electromagnetic cascade in the microwave background and reach Earth in the form of multiple photons of lower energy. The electromagnetic shower in the microwave background subdivides the initial PeV photon energy, leading to multiple photons in the range of GeV to TeV energies by the time the gamma rays arrive at Earth~\cite{Protheroe1993,Ahlers:2010fw}. If the target itself is not transparent to gamma rays, they will also lose energy before leaving the source.

\begin{figure}[ht!]
\centering
\includegraphics[width=.75\linewidth]{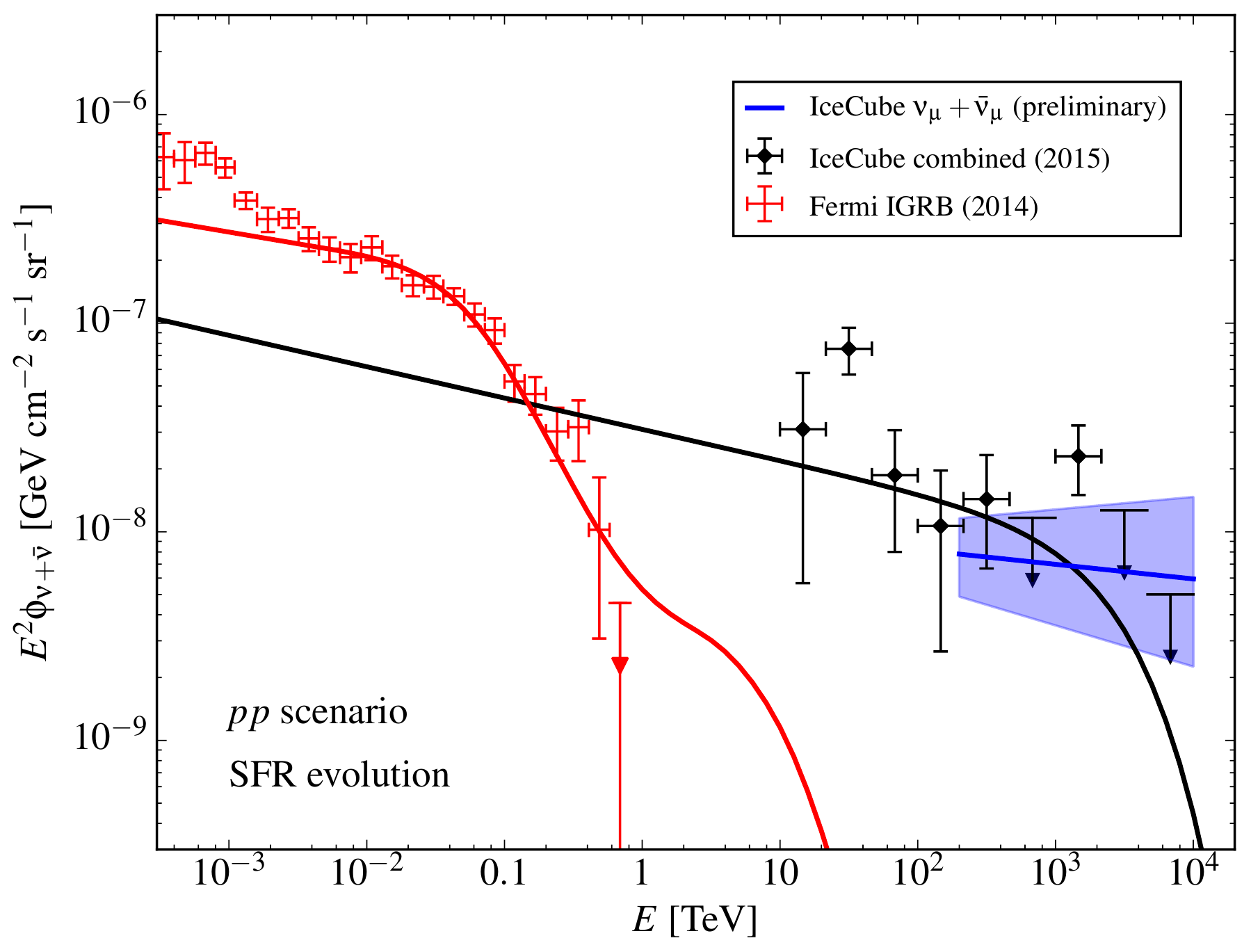}
\caption[]{An early figure that showed that the astrophysical neutrino flux (black line) observed by IceCube qualitatively matches the corresponding cascaded gamma-ray flux (red line) observed by Fermi. We assume that the decay products of neutral and charged pions from $pp$ interactions are responsible for the nonthermal emission in the universe~\protect\cite{Murase:2013rfa}. The black data points are early IceCube results, including the three-year ``high-energy starting event" (HESE) analysis~\protect\cite{Aartsen:2014gkd} and a subsequent analysis lowering the energy threshold for events starting in the detector even further~\protect\cite{Aartsen:2014muf}. Also shown is the best fit to the flux of high-energy muon neutrinos penetrating the Earth. Introducing the cutoff on the high-energy flux, shown in the figure, does not affect the result.}
\label{fig:NewfermiPlot2}
\end{figure}

For illustration, we calculate the gamma-ray flux accompanying the IceCube diffuse cosmic neutrino flux, which is described by a simple power law with spectral index of -2.15 and is consistent with the cosmic neutrino data above an energy of 100~TeV. 
The result, assuming equal multiplicities $\pi^{0}=\pi^{+}=\pi^{-}$, is shown in Fig.~\ref{fig:NewfermiPlot2}. The matching energy densities of the extragalactic gamma-ray flux detected by Fermi and the high-energy neutrino flux measured by IceCube suggest that, rather than detecting some exotic sources, it is more likely that IceCube to a large extent observes the same universe astronomers do. The finding implies that a significant fraction of the energy in the nonthermal universe originates in hadronic processes, indicating a larger role than previously thought. Accordingly, IceCube developed methods, most promisingly real-time multiwavelength observations with astronomical telescopes, to identify the sources and build on the discovery of cosmic neutrinos to launch a new era in astronomy~\cite{Aartsen:2016qbu,Aartsen:2016lmt}.

The overall normalization of the emission has been chosen not to exceed the isotropic gamma-ray background observed by the Fermi satellite (red data). Interestingly, we conclude that the high-energy cosmic neutrino flux above 100~TeV shown in Fig.~\ref{fig:NewfermiPlot2} saturates this limit. The starting event data that extend to lower energies, below 100~TeV, are only marginally consistent with the upper bound (red line). This high intensity of the neutrino flux at lower energies compared to the isotropic gamma-ray background implies that in these dark sources the gamma rays' energy ends up below Fermi's threshold~\cite{Murase:2015xka,Bechtol:2015uqb}. 

Interestingly, the common energy density of photons and neutrinos is also comparable to that of the ultra-high-energy extragalactic cosmic rays~\cite{Aab:2015bza}. The matching energy densities in cosmic rays and neutrinos is suggestive of a common origin.

The extragalactic gamma-ray background observed by Fermi~\cite{Ackermann:2014usa} has contributions from identified sources, mostly blazars, on top of an isotropic gamma-ray background (IGRB) shown in Fig.~\ref{fig:NewfermiPlot2}. Furthermore, this diffuse contribution is expected to result from emission from the same class of gamma-ray sources that are individually below Fermi's point-source detection threshold (see, e.g.,~\cite{DiMauro:2015tfa}). This speculation received further support when IceCube developed methods for performing real-time multiwavelength observations~\cite{Aartsen:2016qbu,Aartsen:2016lmt} that led to the evidence for identification of a distant flaring blazar as a cosmic-ray accelerator in a multimessenger campaign launched by a 290-TeV energy neutrino detected on September 22, 2017~\cite{IceCube:2018dnn}.

\section{The first coincident observations of high-energy neutrinos and gamma rays}

Neutrinos only originate in environments where protons are accelerated to produce pions and other particles that decay into neutrinos. Neutrinos can thus exclusively pinpoint cosmic-ray accelerators, and this is exactly what one neutrino did on September 22, 2017.

IceCube detects muon neutrinos, a flavor that leaves a well-reconstructed track in the detector roughly every five minutes. Most of them are low-energy neutrinos produced in Earth's atmosphere, which are of interest for studying properties of the neutrinos themselves but are a persistent background when doing neutrino astronomy. In 2016, IceCube installed an online filter that, in real time, selects from this sample of more than $10^5$ per year those very high energy neutrinos that are likely to be of cosmic origin~\cite{Aartsen:2016lmt}. The computer cluster located on the ice above the experiment reconstructs their energy and celestial coordinates, typically in less than one minute, and distributes the information automatically via the Gamma-ray Coordinate Network to a group of telescopes around the globe and in space for follow-up observations. These telescopes look for coincident electromagnetic radiation from the arrival direction of the neutrino, creating the opportunity for revealing its origin.  

The tenth such alert~\cite{2017GCN.21916....1K}, IceCube-170922A, on September 22, 2017, reported a well-reconstructed muon neutrino with an energy of 290 TeV and, therefore, with a significant probability of originating in an astronomical source. The Fermi telescope detected a flaring blazar aligned with the cosmic neutrino within 0.06 degrees. The source is a known blazar, a supermassive rotating black hole producing high-energy gamma rays in a jet aligned with its rotation axis that is directed at Earth. The blazar TXS 0506+056 was originally flagged by the Fermi~\cite{2017ATel10791....1T} and Swift~\cite{2017ATel10792....1E} satellite-borne telescopes. Follow-up observations with the MAGIC air Cherenkov telescope~\cite{2017ATel10817....1M} identified it as a relatively rare TeV-energy blazar with the potential to produce the very high energy neutrino detected by IceCube. The source was subsequently scrutinized in X-ray, optical, and radio wavelengths. In total, more than 20 telescopes observed the flaring blazar as a highly variable source in a high state~\cite{IceCube:2018dnn}.

TXS 0506+056 had been relatively poorly studied until now, although it had been identified as the preeminent gamma-ray source in the EGRET sky among those sources that produced two photons or more with energies above 40 GeV \cite{Dingus:2001hz}. Observations triggered by the neutrino alert yielded a treasure trove of multiwavelength data. An optical telescope eventually measured its distance~\cite{Paiano:2018qeq}, which corresponds to a redshift of 0.34. Its distance of 1.7\,Gpc points to a special galaxy, which sets it apart from the ten-times-closer blazars, such as the Markarian sources that dominate the extreme gamma-ray sky observed by Fermi. Getting all the elements of this multimessenger observation together has been challenging. The hadronic contribution to the high-energy emission is constrained by the X-ray flux expected in the routine modeling of blazars. In the end, the conclusion is that, in conventional blazar models, a positive fluctuation is required to accommodate the single neutrino in 2017---never mind the dozen in the 2014-15 neutrino flare that we will discuss later. For details, see \cite{Keivani:2018rnh, Gao:2018mnu, Cerruti:2018tmc, Zhang:2018xrr, Gokus:2018lgx, Sahakyan:2018voh, Murase:2018iyl}.

It is important to realize that nearby blazars like the Markarian sources are at a redshift that is ten times smaller, and therefore TXS 0506+056, with a similar flux despite the greater distance, is one of the most luminous sources in the universe. It likely belongs to a special class of blazars that accelerate proton beams, as revealed by the neutrino. This explains the fact that a variety of previous attempts to associate the arrival directions of cosmic neutrinos with the bulk of Fermi blazars~\cite{Aartsen:2016lir} have failed. More about that later.

Informed by the multimessenger campaign, IceCube searched its 9.5-year archival neutrino data up to and including October 2017 in the direction of IC-170922A using a likelihood search routinely used in previous searches for time-varying sources. This revealed a spectacular burst in the 2014-2015 data of over a dozen high-energy neutrinos in 110 days, 19 on a background of less than 6, with a spectral index similar to the one observed for the diffuse cosmic neutrino spectrum~\cite{IceCube:2018cha}.

\section{Flaring sources and the total high-energy neutrino flux}

In this section we make two more points regarding the TXS 0506+056 neutrino source~\cite{Halzen:2018iak}:
\begin{itemize}[leftmargin=0.15 in]
\item If every blazar produced neutrinos at the level of  TXS 0506+056, the sources would overproduce the total flux observed by IceCube by a factor of 20. TXS 0506+056 must indeed belong to a special subclass of sources, as already suggested by the large redshift.
\item A source that produces 13 neutrinos in 4 months has a target density for producing neutrinos that is large and therefore opaque to high-energy gamma rays. It takes a massive accretion event onto the black hole to accommodate the 2014-15 observation. The 2014-15 burst cannot be, and is not, accompanied by a large Fermi flare.
\end{itemize}

These conclusions are derived below. If the reader is not interested in the details, this section can be skipped; in the following section, we will present intriguing evidence that the accretion is associated with galaxy mergers, with the accreting matter or radiation in galaxy mergers providing the target material for producing the neutrinos that is missing in the vanilla one-zone blazar scenario. We will focus on the single neutrino flare in 2014-15 that dominates the total flux of the source over the 9.5 years of IceCube observations, leaving IC-170922A as a less luminous second flare in the data sample.

Next we will study the relation between the TXS 0506+056 flux and the all-sky neutrino flux and show that a subset of about 5\% of all blazars, bursting once in 10 years at the level of TXS 0506+056 in 2014, can accommodate the total neutrino flux observed by IceCube. The total flux of high-energy neutrinos from a fraction ($\mathcal{F}$) of a class of sources with density $\rho$ in the universe, neutrino luminosity $L_\nu$,  and episodic emission of flares of duration $\Delta t$ over a total observation time $T$ is
\begin{eqnarray}
\sum_{\alpha} E_\nu^2 \frac{d N_\nu}{dE_\nu} = \frac{1}{4\pi} \frac{c}{H_0} \xi_z L_\nu \rho \mathcal{F} \frac{\Delta t}{T} \,,
\end{eqnarray}
where $\xi_z$ is a factor of order unity that parametrizes the integration over the redshift evolution of the sources \cite{Halzen:2018iak}.
Applying this relation to the 2014-15 TXS 0506+056 burst, which dominates the fluency from the sources over the 9.5 years of neutrino observations, yields
\begin{eqnarray}\label{diffuse_flux}
\begin{aligned}
3\times10^{-11}\, {\rm{TeV cm^{-2} s^{-1} sr^{-1}}} = &\frac{ \mathcal{F}}{4\pi}  \frac{c}{H_0} \bigg(\frac{\xi_z}{0.7 \rm{}} \bigg) \bigg(\frac{L_\nu}{1.2\times10^{47}\, \rm{erg/s}} \bigg)\\ &\bigg(\frac{\rho}{1.5\times10^{-8}\, \rm{Mpc^{-3}}} \bigg) \bigg(\frac{\Delta t}{110 \,{\rm d}} \frac{10 {\, \rm yr}}{T}\bigg)\,, 
\end{aligned}
\end{eqnarray}
a relation which is satisfied for $\mathcal{F}\sim0.05$. In summary, a special class of blazars that undergo $\sim110$-day duration flares like TXS 0506+056 once every 10 years accommodates the observed diffuse flux of high-energy cosmic neutrinos. The class of such neutrino-flaring sources represents 5\% of all sources. The argument implies the observation of roughly 120 neutrinos per year of muon flavor. This is indeed the flux of cosmic neutrinos that corresponds to the $E^{-1.9}$ diffuse flux. (Note that the majority of these neutrinos cannot be separated from the atmospheric background, leaving us with the reduced number of very high energy events discussed in the previous sections). 

As previously discussed, the energetics of the cosmic neutrinos is matched by the energy of the highest energy cosmic rays. Their energy densities are related by
\begin{eqnarray}
\frac{1}{3}\sum_{\alpha} E_\nu^2 \frac{d N_\nu}{dE_\nu} \simeq \frac{c}{4 \pi}\,\bigg( \frac{1}{2}
(1-e^{-\tau})\, \xi_z t_H \frac{dE}{dt} \bigg)\,.
\end{eqnarray}
The cosmic rays' injection rate $dE/dt$ above $10^{16}$ eV is $(1-2) \times 10^{44}\,\rm erg$\, $\rm Mpc^{-3}\,yr^{-1}$~\cite{Ahlers:2012rz, Katz:2013ooa}. From this and the relation above, it follows that the energy densities match for an opacity of the source of $\tau \gtrsim 0.4$. This high-opacity requirement is consistent with the premise that a special class of efficient sources is responsible for producing the high-energy cosmic neutrino flux seen by IceCube. The sources must contain sufficient target density in photons, possibly protons, to generate a value of $\tau$ of order unity. It is clear that the emission of flares producing the large number of cosmic neutrinos detected in the 2014 burst must correspond to a major accretion event onto the black hole lasting a few months. The pionic photons will lose energy in the opaque source and the neutrino emission is not accompanied by a gamma-ray flare, as was the case for the 2017 event; see~\cite{Halzen:2018iak} for details. The Fermi data, consistent with the scenario proposed, reveal photons with energies of tens of GeV but no dramatic enhancement in gamma-ray activity~\cite{Garrappa:2018}.

In the modeling of hadronic blazars, it is typically assumed that the interaction of accelerated protons with photons is responsible for generating pions that decay into neutrinos. The interaction is dominated by the $\Delta$-resonance, with $p \gamma \rightarrow \Delta \rightarrow \pi N$. Accelerated cosmic rays may interact with photons in the jet or with stationary photons provided by the accretion disk.  Assuming this for illustration, the required efficiency of the source to produce pions, given by $(1-e^{-\tau}) \simeq \tau$ and often referred to as $f_\pi$, depends on the target photons' energy, the Lorentz factor of the jet, the luminosity of the photons, and the duration of the flare: 
\begin{eqnarray}
\tau \simeq
\frac{L_{\gamma}}{E_{\gamma}}\frac{1}{\Gamma^2 \Delta t} \frac{3
\sigma_\Delta \langle x_{p \rightarrow \pi} \rangle}{4\pi c^2}.
\end{eqnarray}
Since TXS 0506+056 is an intermediate synchrotron-peaked blazar, we consider the target photon at the synchrotron energy peak, 10 eV, to constrain the parameter space for the pion efficiency:
\begin{eqnarray}
\begin{aligned}
\tau \gtrsim  0.4 \simeq
\bigg(\frac{L_{\gamma}}{2\times 10^{46}\, \rm{erg/s}}\bigg)
\bigg(\frac{10\, \rm eV}{E_{\gamma}}\bigg)
\bigg(\frac{1}{\Gamma^2} \bigg) \\
\bigg(\frac{110 \, \rm d}{\Delta t} \bigg)
\bigg(\frac{3 \sigma_\Delta \langle x_{p \rightarrow \pi} \rangle}{4\pi c^2} \bigg).
\end{aligned}
\end{eqnarray}
This suggests that sources with a small Lorentz factor and UV luminosity of $\mathcal O(10^{46})$ can satisfy the requirements for producing the high-energy cosmic neutrinos observed. This level of UV luminosity is compatible with the {\em Swift}-UVOT measurement reported by \cite{Padovani:2018acg}. Similar values for UV luminosity were also reported in \cite{Krauss:2014tna}. In addition, the study of the radio structure of TXS 0506+056 suggests a small Lorentz factor \cite{Kun:2018zin}. The case for a small Lorentz factor also follows arguments made in the context of a tentative observation of high-energy neutrinos by AMANDA in coincidence with a flare of 1ES 1959+560 in 2002~\cite{Halzen:2005pz}.

A key question is whether the neutrino and gamma-ray spectra for the 2014 neutrino burst from TXS 0506+056 satisfy the multimessenger relationship introduced in Section \ref{IC_Fermi}. With the low statistics of the very high energy gamma-ray measurements during the burst period, the energetics is a more robust measure for evaluating the connection, especially because the source is opaque to high-energy gamma rays, as indicated by the large value of $\tau$, and the pionic gamma rays will lose energy inside the source before cascading in the microwave photon background; for details, see \cite{Halzen:2018iak}. 

The gamma-ray opacity can be related to $\tau$ by 
\begin{equation}
\tau _ { \gamma \gamma } \approx \frac { \eta _ { \gamma \gamma } \sigma _ { \gamma \gamma } } { \eta _ { p \gamma } \hat { \sigma } _ { p \gamma }} \tau\,,
\end{equation}
where $\hat {\sigma} _ {p\gamma} \sim 0.7 \times 10 ^{-28} \mathrm {cm}^{2}$, $\sigma_{\gamma\gamma} \simeq 6.65 \times 10 ^ { - 25 } \mathrm{cm}^2$, and the threshold factors are given by $\eta _ { \gamma \gamma } \sim 0.1$ and $\eta _ { p \gamma } \simeq 1$ ~\cite{Murase:2015xka}.  This implies $\tau _ {\gamma \gamma } \simeq \mathcal{O}(100)$ for $\tau \gtrsim 0.4$ . This high opacity makes it impossible for the very high energy pionic gamma rays, with energies similar to those of the neutrinos, to leave the source.

Absorption and interactions intrinsic to the source due to the high opacity, followed by the interaction with EBL, will result in a suppressed gamma-ray flux, and hence a gamma-ray flare is not expected when the source is a highly efficient neutrino emitter \cite{Halzen:2018iak}. This is consistent with the Fermi observation of the source during the neutrino flare in 2014 \cite{Padovani:2018acg, Garrappa:2018}. Hardening of the spectrum reported by \cite{Padovani:2018acg} can be explained in this picture; see \cite{Halzen:2018iak}. The absence of gamma rays coincident with a neutrino burst could also be a result of secondary gamma rays spreading in a larger opening angle compared to the original beam; see \cite{Neronov:2002xv} for details.

It is worth noting that this model for the diffuse neutrino flux clarifies why earlier attempts to associate it with the bulk of blazars were unsuccessful and why the limits built on source population and nondetection of a steady source do not apply. First, the time-integrated studies are not applicable to time-dependent sources. Moreover, with a yet to be completely identified subclass of energetic sources with lower density in the universe responsible for the total flux, the constraints on blazars obtained from the relation between the point source limits and the diffuse flux are mitigated.

\section{Closing in on the sources?}

One straightforward way that a subclass of blazars can stand out is by redshift evolution: powerful proton accelerators producing neutrinos may have been active in the past but are no longer active today. This would explain the relatively large redshift of TXS 0506+056~\cite{Neronov:2018wuo}; it is the closest source from a set of sources that only accelerated cosmic rays at early redshifts. Another possibility considers focusing on the requirement of a large target density in photons or protons to produce neutrinos at the level observed from TXS 0506+056. Importantly, follow-up studies of the source with the high-resolution VLBA array have revealed a cosmic jet collision, potentially a collision of two jets on pc-scales~\cite{Britzen2019}. 
Abundant neutrino emission will result from the interaction of jetted material. This speculation has been reinforced by the recent observation of the blazar PKS 1502+106 in coincidence with the recent 300-TeV neutrino alert IC-190730~\cite{2019ATel12971....1L}. OVRA radio observation~\cite{2019ATel12996....1K} indicates that the neutrino is coincident with the highest flux density of a flare at 15\,GHz that started five years ago~\cite{2016A&A...586A..60K}. This matches the similar long-term radio outburst seen from TXS 0506+056 and may be a signature of a merger event.

There is a development, possibly related, associated with the results of the time-integrated search for neutrino sources that recently delivered a neutrino sky map for 10 years of data ~\cite{Carver:2019jcd}. There is evidence at the $3\sigma$ level that the map is no longer isotropic. The anisotropy is contributed by four sources that show evidence for clustering at the $4\sigma$ level, pretrial. The strongest of these sources is the nearby Seyfert galaxy NGC 1068 (also known as Messier 77); the second is TXS 0506+056, despite its strong episodic emission. NGC 1068 is an active galaxy with starburst activity. There is evidence for shocks near the core and for molecular clouds with densities of more than $10^5\,cm^{-3}$. Its origin is a merger onto the black hole, either with a satellite galaxy or, more likely, with a star-forming region ~\cite{2014A&A...567A.125G}.

A cursory review of the theoretical literature on the production of neutrinos in galaxy mergers is sufficient to conclude that it can accommodate the observations of both the individual sources discussed above as well as the total flux of cosmic neutrinos ~\cite{Kashiyama:2014rza,Yuan:2017dle,Yuan:2018erh}.

\section{Summary}

These speculations are qualitative and may be premature, but the hope is that multimessenger astronomy will provide us with more clues after the breakthrough event of September 22, 2017, which generated an unmatched data sample over all wavelengths of the electromagnetic spectrum from a cosmic-ray accelerator.

Unlike the previous SN1987A and GW170817 multimessenger events, this event could not have been observed with a single instrument. Without the initial coincident observation, IC-170922A would be just one more of the few hundred high energy cosmic neutrinos detected by IceCube and the accompanying radiation just one more flaring blazar observed by Fermi-LAT. Neutrino astronomy was born with a supernova in 1987. Thirty years later, this recent event involves neutrinos that are tens of millions of times more energetic and are from a source a hundred thousand times more distant.

\section{Acknowledgments}

I thank Ali Kheirandish for generously contributing to the preparation of this talk. Discussion with collaborators inside and outside the IceCube Collaboration, too many to be listed, have also greatly shaped this presentation. Thanks. This work was supported in part by the U.S. National Science Foundation under grants~PLR-1600823 and PHY-1607644 and by the University of Wisconsin Research Committee with funds granted by the Wisconsin Alumni Research Foundation.

\bibliographystyle{ICRC}
\bibliography{bib}

%

\end{document}